\documentclass[sigconf]{acmart}

\DeclareUnicodeCharacter{2064}{}
\AtBeginDocument{%
  \providecommand\BibTeX{{%
    \normalfont B\kern-0.5em{\scshape i\kern-0.25em b}\kern-0.8em\TeX}}}

\copyrightyear{2024}
\acmYear{2024}
\setcopyright{acmlicensed}\acmConference[NordiCHI 2024]{Nordic Conference on Human-Computer Interaction}{October 13--16, 2024}{Uppsala, Sweden}
\acmBooktitle{Nordic Conference on Human-Computer Interaction (NordiCHI 2024), October 13--16, 2024, Uppsala, Sweden}
\acmDOI{10.1145/3679318.3685349}
\acmISBN{979-8-4007-0966-1/24/10}

\acmConference[NordiCHI '24]{Nordic Human-Computer Interaction Conference}{October 13--16, 2024}{Uppsala, Sweden}

\usepackage{todonotes}
\let\xtodo\todo
\renewcommand{\todo}[1]{\xtodo[inline,color=orange!75]{#1}}

\usepackage{xcolor}
\usepackage{subcaption}
\usepackage{fontawesome}

\definecolor{lightgray}{gray}{0.7}

\begin{document}
\title[Comparing Visual, Haptic, and Visuohaptic Encoding on Memory Retention]{Comparing the Effects of Visual, Haptic, and Visuohaptic Encoding on Memory Retention of Digital Objects in Virtual Reality}
 
\author{Lucas Siqueira Rodrigues}
\email{lucas.siqueira.rodrigues@hu-berlin.de}
\orcid{0000-0001-7675-136X}
\affiliation{%
  \institution{HU Berlin}
  \city{Berlin}
  \country{Germany}
}

\author{Timo Torsten Schmidt}
\email{timo.t.schmidt@fu-berlin.de}
\orcid{0000-0003-1612-1301}
\affiliation{%
  \institution{FU Berlin}
  \city{Berlin}
  \country{Germany}
}

\author{Johann Habakuk Israel}
\email{israel@htw-berlin.de}
\orcid{0000-0002-8513-6892}
\affiliation{%
  \institution{HTW Berlin}
  \city{Berlin}
  \country{Germany}
}

\author{Stefan Zachow}
\email{zachow@zib.de}
\orcid{0000-0001-7964-3049}
\affiliation{%
  \institution{Zuse Institute Berlin}
  \city{Berlin}
  \country{Germany}
}

\author{John Nyakatura}
\email{john.nyakatura@hu-berlin.de}
\orcid{0000-0001-8088-8684}
\affiliation{%
  \institution{HU Berlin}
  \city{Berlin}
  \country{Germany}
}

\author{Thomas Kosch}
\email{thomas.kosch@hu-berlin.de}
\orcid{0000-0001-6300-9035}
\affiliation{%
  \institution{HU Berlin}
  \city{Berlin}
  \country{Germany}
}

\renewcommand{\shortauthors}{Siqueira Rodrigues et al.}

\begin{abstract}

Although Virtual Reality (VR) has undoubtedly improved human interaction with 3D data, users still face difficulties retaining important details of complex digital objects in preparation for physical tasks. To address this issue, we evaluated the potential of visuohaptic integration to improve the memorability of virtual objects in immersive visualizations. In a user study (N=20), participants performed a delayed match-to-sample task where they memorized stimuli of visual, haptic, or visuohaptic encoding conditions. We assessed performance differences between these encoding modalities through error rates and response times. We found that visuohaptic encoding significantly improved memorization accuracy compared to unimodal visual and haptic conditions. Our analysis indicates that integrating haptics into immersive visualizations enhances the memorability of digital objects. We discuss its implications for the optimal encoding design in VR applications that assist professionals who need to memorize and recall virtual objects in their daily work.

\end{abstract}


\begin{teaserfigure}
  \includegraphics[width=\textwidth]{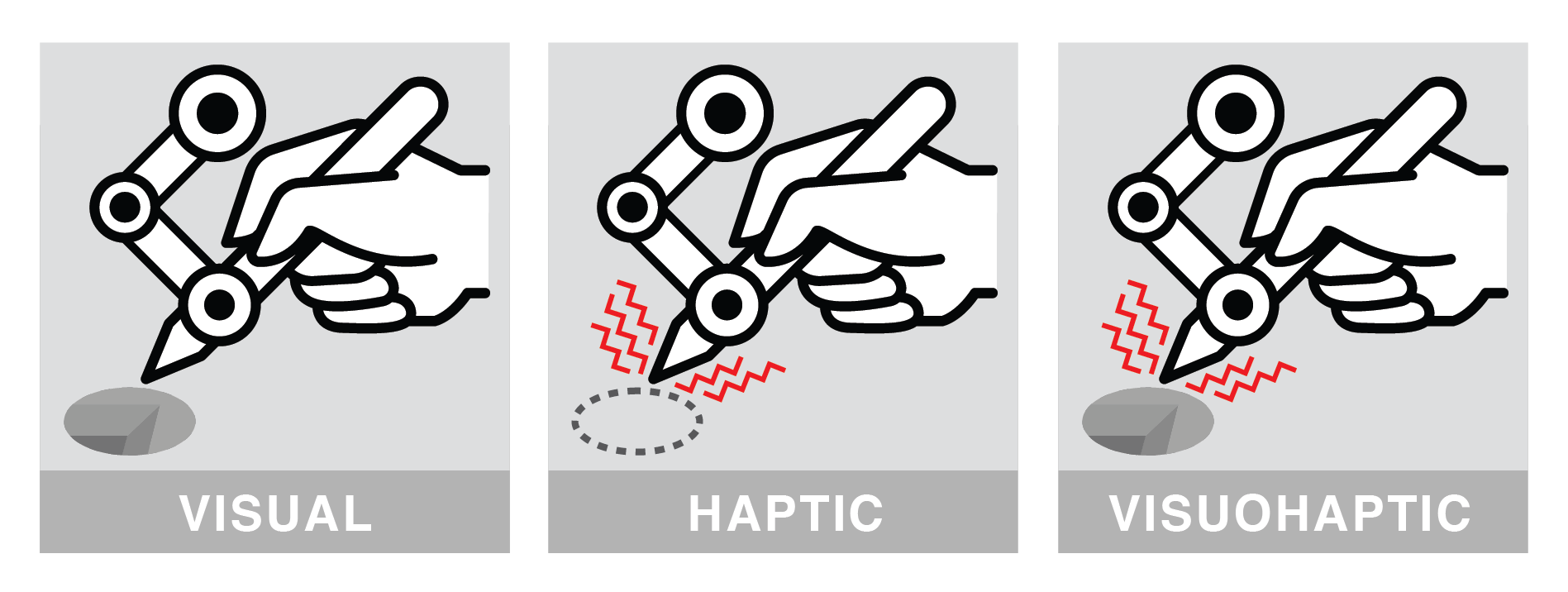}
  \caption{This paper evaluates the visual, haptic, and visuohaptic encoding conditions on the memorization of objects in Virtual Reality. Our results show that visuohaptic encoding produces the lowest error rate and response time.}
  \Description{This paper evaluates the effects of visual, haptic, and visuohaptic encoding conditions on memorizing objects in Virtual Reality. Our results show that visuohaptic encoding produces the lowest error rate and response time.}
  \label{fig:teaser}
\end{teaserfigure}



\maketitle

\section{Introduction}
Virtual Reality (VR) systems have predominantly relied on vision as their primary sensory modality for user interaction, which is particularly the case for immersive scientific visualization platforms facilitating the exploration of digital objects~\cite{moreau_visual_2013, ghinea_multisensory_2012, kraus_value_2021}. For example, professionals working in the fields of medicine~\cite{desselle_augmented_2020, meier_virtual_2001} and paleontology~\cite{cunningham_virtual_2014, rodrigues_immersive_2022} leverage such systems to examine digital replicas in preparation for real-world interventions. Despite immersive visualization advancements, challenges arise when users must memorize complex details of working objects and retain such information across contexts to process real-world objects. Human errors often occur when users rely on visual memory to recall important object features once they have left the VR environment to perform a real-world task~\cite{lamers_changing_2021, roo_understanding_2018}. 

To reduce errors from visual memory recalls, the integration of haptic feedback into VR settings is regarded as a potential improvement to the interaction with digital models~\cite{lacey_multisensory_2011, klatzky_identifying_1985, tokuda_effectiveness_2009, wolf_visuo-haptic_2022, rydstrom_effect_nodate}. Visuohaptic integration, the coherent combination of vision and touch, has been investigated for its potential to enhance cognitive tasks related to perceiving and manipulating objects within virtual environments~\cite{kaiser_visuo-haptic_2010, lalanne_crossmodal_2004, kaas_effect_2007}. The well-documented cognitive synergy between touch and sight suggests that this approach could lead to more robust mental representations of digital objects \cite{easton_vision_1997, amedi_visuo-haptic_2001, james_haptic_2002, lacey_vision_2007}.

Although the literature documents certain cognitive benefits of immersive visuohaptic object exploration, a research gap remains regarding understanding the memorability of such interactions. While previous studies have primarily focused on different aspects of sensory processing, attention, and perception, the impact of visuohaptic encoding on the retention of object characteristics is not sufficiently understood and requires further investigation~\cite{spence_crossmodal_2004, grunwald_human_2008}.

Here, we present a behavioral study to investigate whether visuohaptic interaction in VR enhances user performance in a delayed match-to-sample task (DMTS), a well-established procedure for assessing the accuracy of object retention over time~\cite{grefkes_crossmodal_2002, easton_vision_1997}. The DMTS task is commonly applied in psychology and neuroscience to investigate the cognitive processes involved in the retention of visual \cite{romo_neuronal_1999, harrison_decoding_2009} and haptic information \cite{schmidt_content-specific_2017, schmidt_brain_2018, wu_overlapping_2018}. Participants retained stimuli encoded by visual, haptic, or visuohaptic exploration and, after a delay, performed a two-alternative forced choice (2AFC) to distinguish the retained stimulus from a foil. In this within-subject design study, we tested for differences between the three experimental conditions regarding error rates and response times.

Our data indicated that supplementing haptic information significantly reduced error rates, as we found improved performance in the visuohaptic condition compared to the visual or haptic-only conditions. However, visuohaptic encoding did not reduce response time. Our findings are relevant to the design of human-computer interaction interfaces that optimally inform users of the properties of complex digital objects in a manner that aligns with natural human sensory integration capabilities. Leveraging such efficacy might improve the design of interfaces targeting practitioners whose workflows require accurate memorization of digitalized objects within VR environments. 

\section{Related Work}
\subsection{Virtual Reality in Data Visualization}
VR is becoming increasingly common in visualizing 3D models and image datasets \cite{besancon_state_2021, korkut_visualization_2023}. Since the late 80s, researchers have been assessing the utility of VR in enhancing the exploration and analysis of complex data for various applications, including visualizing heritage artifacts and planning surgical interventions \cite{miller_good_1995, reilly_visualizing_1992, simpson_immersive_2000, cruz-neira_surround-screen_1993, rodrigues_voxsculpt_2023}. Recently, VR has been integrated into commercially available and open-source scientific visualization and analysis platforms such as Slicer, Amira, and ParaView \cite{pinter_slicervr_2020, stalling_amira_2005, shetty_immersive_2011}. This trend may be explained as an attempt to leverage the technology's known capabilities in improving users' ability to interpret and manipulate complex object structures, given the rich set of spatial and depth cues provided by immersive data exploration \cite{bryson_virtual_1996, pausch_quantifying_1997, mccormick_visualization_1988}. Subsequently, VR has been demonstrated to increase \textit{presence}, the sensation of being immersed in a virtual environment that can extend to responding to digital stimuli as if they were real \cite{heeter_being_1992}. In its turn, presence may be valuable to scientific visualization as it knowingly enhances performance on spatial cognition tasks, improving situational awareness, spatial judgments, and navigation \cite{maneuvrier_presence_2020, pausch_quantifying_1997, simpson_immersive_2000, palombi_role_2023}. In essence, VR is becoming increasingly common in visualizing 3D models and image datasets due to its potential to enhance interaction with complex data representations and foster cognitive performance in scientific visualization tasks.

\subsection{Memorability in Immersive Visualizations}
While the literature presents a number of cognitive advantages of immersive visualization, studies assessing its effect on memorability yielded mixed results, indicating that its potential retention improvements may be context-dependent and limited to certain workload conditions \cite{hamilton_immersive_2021, smith_immersion_2021, brooks_specificity_1999, 10.1145/3582272}. Under the premise that both encoding and testing occur under the same immersive conditions, Krokos et al. suggested that VR's embodied interaction with virtual objects could yield superior information recall in comparison with desktop displays \cite{krokos_virtual_2019}. Buttussi and Chittaro observed improved retention of spatial knowledge regarding distance estimations and spatial relations with VR compared to a touchscreen device \cite{buttussi_acquisition_2023}. Wrage et al. compared tracked and untracked head-mounted display rotations and attributed improved object location recall performance in VR to self-initiated vestibular and proprioceptive rotations, demonstrating that the advantages of VR in encoding spatial information may be extraocular \cite{wraga_spatial_2004}. However, immersive learning seems to be subject to the \textit{context-dependent memory} effect, which posits that memory performance depends on whether encoding and recall occur in matching environments \cite{godden_context-dependent_1975}. This effect was initially demonstrated between incongruent physical environments. Still, Lamers and Lanen replicated this study to assess its applicability to immersion and found that switching from a virtual to a physical environment impacts the ability to recall information \cite{lamers_changing_2021}. Roo et al. also observed that users cannot transfer object information between virtual and physical spaces and found higher error rates in position estimates learned in VR \cite{roo_understanding_2018}. Shin et al. also demonstrated that object memory is better recalled when encoding and retrieval contexts are congruent \cite{shin_context-dependent_2021}. An increased sense of presence and its associated attention arousal could also be responsible for lower performance in subsequent real-world memory tasks, as demonstrated by Bailey et al. through free and cued recall tasks \cite{bailey_presence_2012}. In summary, immersion improves visualization memorability as long as encoding and retrieval occur within the virtual environment. Still, this effect might be context-dependent as users appear to be unable to retain object information optimally between virtual and physical settings. 

\subsection{Improving Immersive Visualization Memorability with Haptics}
As VR visualizations endure memorability issues, it is worthwhile to investigate further factors that may improve the retention of objects displayed in immersive environments. Studies comparing immersive visualizations and data physicalization might point to haptics as a candidate for enhancing memorability in this context. Ren and Hornecker compared memorability in VR and tangible modalities and found that VR had inferior recollection and understanding performance compared to data physicalization, an approach involving physical representations of information \cite{ren_comparing_2021}. Jansen et al. conducted a similar study that attributed this effect to the sense of touch, highlighting the importance of haptics in cognitive processes \cite{jansen_evaluating_2013}. Integrating information from vision and touch may result in perceptual advantages, including increased feature salience, resolution of perceptual ambiguities, and unified object perception \cite{lalanne_crossmodal_2004}, thereby allowing the formation of a more detailed and accurate mental image of objects. Beyond research on visualization, findings in the domain of cognitive neuroscience also raise the question of whether integrating sensory information from multiple modalities could potentially improve retention. While visual and haptic information is initially processed in distinct brain regions, the modality-specific processing ultimately converges in brain regions that code modality-overarching abstract mental representations of shape \cite{easton_transfer_1997, easton_vision_1997, craddock_size-sensitive_2009}. Neuroimaging studies have demonstrated specified brain regions that integrate haptic and visual shape and object information \cite{amedi_visuo-haptic_2001, grunwald_human_2008, james_haptic_2002}. Supplementing haptic information might, therefore, support the accuracy of a formed mental representation of an object.  Behavioral studies involving object identification indicate that vision and haptics share representations \cite{loomis_analysis_1982, klatzky_identifying_1985}. These sensory modalities rely similarly on shape information and exhibit analogous object identification error patterns \cite{gaissert_categorizing_2012}. While comparisons between VR visualization and data physicalizations hint that haptics could enhance visualization memorability, extensive research on the cognitive synergy between touch and sight upholds the value of investigating this proposition due to its potential to produce robust mental representations of digital objects.

\subsection{Haptic Integration into Data Visualization}
The integration of haptics into visualizations has been explored since the early 1990s when Brooks et al. presented a haptic display that implemented force fields in a virtual environment that allowed the interaction with protein structures \cite{brooks_project_1990}. Iwata and Noma later introduced the concept of \textit{volume haptization} \cite{iwata_volume_1993}, and Avila and Sobierajski presented a haptic interactive method for displaying tomographic volume data as force feedback using Massie and Salisbury's PHANToM device \cite{avila_haptic_1996, massie_phantom_1994}. Haptic rendering developments such as Ruspini and Khatib's finger-proxy algorithm \cite{ruspini_haptic_1997} and Zilles and Salisbury's god-object method \cite{zilles_constraint-based_1995} addressed their technical challenges, and established haptics in data visualization with proven benefits. For instance, interfaces combining haptic and visual rendering have facilitated understanding the complex scalar, vector, and tensor fields \cite{lawrence_synergistic_2004}. As occlusion is a common issue in volume rendering, haptics have been proven to improve the detection of faint data structures \cite{palmerius_impact_2009}. The introduction of haptics in this context has also been established to improve target selection \cite{wall_quantification_2000}, path following \cite{passmore_effects_2001, shen_direct_2019}, and volumetric data navigation \cite{mendez_haptic-assisted_2005}. The integration of haptics in visualizations has been demonstrated to enhance conceptual learning \cite{bivall_haptic_2011, bara_haptics_2007}. Haptics also have a demonstrated ability to improve presence in immersive visualizations \cite{kreimeier_evaluation_2019}. Despite the benefits of integrating haptics into immersive visualizations, potential improvements for the memorability of digital objects have been studied to a far lesser extent.

\subsection{Visuohaptic Integration Effects in Object Retention}
Regarding the effects of visuohaptic integration on object memory retention, several behavioral studies have investigated the impact of combining vision and active touch on object retention and working memory. For example, Desmarais et al. explored the impact of visuohaptic integration on the identification of physical objects, focusing on the influences of stimulus similarity, cross-modality transfer and interference, and congruence to demonstrate that haptic and visual identification rely on shared representations \cite{desmarais_visuo-haptic_2017}. Comparing the support of vibrotactile and exoskeleton force feedback gloves in identifying a limited set of familiar primitive shapes, Kreimeier et al. reported that vibrotactile feedback increased detection rates in comparison with force feedback but negatively impacted speed \cite{kreimeier_evaluation_2019}. Kalenine et al. demonstrated that active touch combined with visual cues improved shape identification of physical objects compared to visual learning alone \cite{kalenine_visual_2011}.  Using a pattern-matching task to investigate the interplay of visual and haptic processing in cognitive load, Seaborn and colleagues found that the incorporation of fingertip vibrotactile cues in this task does not hamper working memory as previously hypothesized, adding that integrating visual and haptic information improves recall performance compared to unimodal presentation \cite{seaborn_exploring_2010}. Exploring the effects of visual, haptic, and visuohaptic exploration as prior knowledge to visual learning to test its impact on learning speed and recognition performance, Jüttner et al. found significant effects of prior haptic knowledge in measured outcomes \cite{juttner_when_2001}. Similarly, Wijntjes et al. found that combining haptic and visual information led to a more accurate perception of 3D shapes compared to vision alone, suggesting that touch helps resolve visual ambiguities and enables veridical interpretation of retinal projections \cite{wijntjes_haptic_2009}. 

\vspace{10pt}
\noindent Although prior research has separately improved our understanding of different interactions between immersion and haptics with object memorability, the literature still has a gap in directly examining the potential of visuohaptic integration in enhancing the retention of objects encoded in VR. This study aims to address this literature gap through a behavioral research that compares the effects of different sensory modality conditions in improving accuracy and efficiency in a working memory task, as described in the following section.

\section{Methodology}
We employed a DMTS task in a within-subjects design study with three encoding conditions: haptic, visual, and visuohaptic. The visual presentation was delivered with a VR setup, which was supplemented with a grounded force-feedback haptic device. We assessed error rate and response time to evaluate potential performance advantages due to the supplementation of haptic feedback. Based on the literature reviewed above, we hypothesized the following:

\begin{itemize}
    \item[\textbf{H1:}] The error rate is the lowest in the visuohaptic encoding condition, followed by visual and haptic encoding. 
    \item[\textbf{H2:}] The response time is the lowest in the visuohaptic encoding condition, followed by visual and haptic encoding.
\end{itemize}

\begin{figure}[!ht]
\centering
\includegraphics[width=\columnwidth]{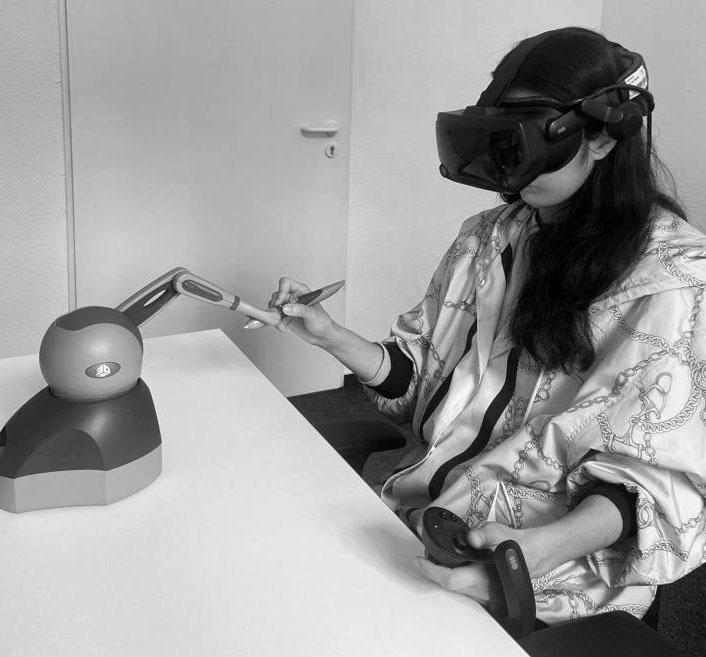}
\caption{A participant using VR and a force-feedback device during an experimental trial.} \label{participant_photo}
\end{figure}
\subsection{Participants}

Participants were right-handed, without report of neurological or psychiatric disorder, and possessed normal or corrected-to-normal vision. None of the participants had prior experience with grounded force-feedback haptic devices. Participants were naive concerning our research objectives. We recruited participants through advertisements posted on online forums. All the participants provided written informed consent. Participants were was compensated monetarily for their time. To ensure data quality, we excluded datasets of participants whose overall performance was below 60\% (fewer than 54 out of 90 correct responses), which represents the minimum score required to perform significantly above chance at $p < .05$ according to a binomial probability distribution. Twenty-three participants completed the study, but three datasets were excluded (48.9\%, 58.1\%, and 45.1\% mean performance), leaving N=20 (age: 31.7 $\pm$ 5.06 years, 8 males, two non-binary participants) for the analyses. Although our sample size is within the typical range of other HCI studies \cite{caine_local_2016}, we acknowledge its limitations in terms of statistical power.

\subsection{Procedure}

\begin{figure*}[!ht]
    \centering
    \includegraphics[width=\textwidth]{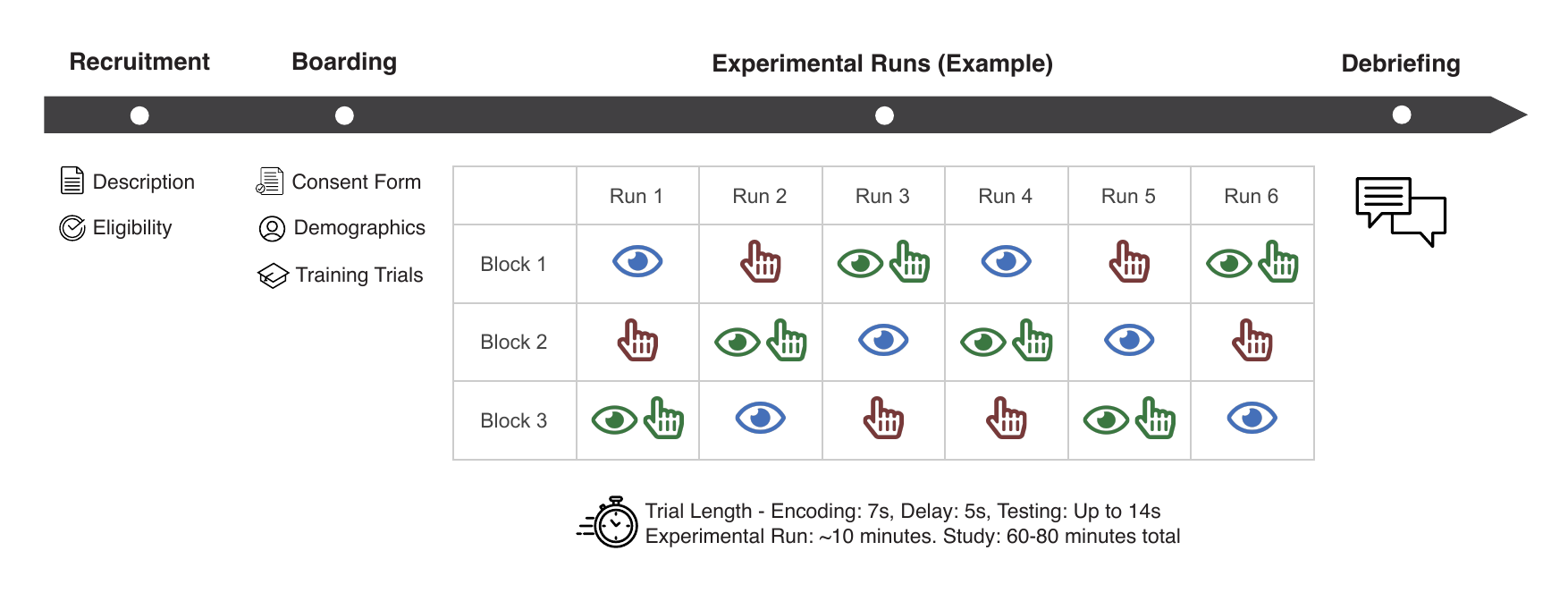}
    \caption[Illustration of the study procedure.]{Illustration of the study procedure. After providing written informed consent and demographics, participants were trained on the task and accommodated with the VR environment and haptic feedback. Participants completed six trial runs containing five trials grouped in single-condition blocks, where the order of conditions was randomized across participants. Cells with {\color[HTML]{4570ba} \faEye} denote the visual condition, while {\color[HTML]{773939} \faHandPointerO}  describe haptic and {\color[HTML]{3b7741} \faHandPointerO\faEye}  are visuohaptic conditions.}
    \label{fig:procedure}
\end{figure*}

Upon arrival, participants provided their written informed consent and demographic information. Next, they were familiarized with the experimental setup, and we consistently aligned the haptic device with their right shoulders to maintain similar pivoting ranges across participants. Then, participants performed nine guided training trials of unlimited time, three for each condition, with manual transitions between the memorizing and testing phases. Participants were allowed to train until they reported feeling properly prepared to perform the task. They were allowed pauses between the runs at their convenience. Every participant completed 90 overall trials, divided into six experimental runs of 15 trials each, taking approximately 10 minutes each, followed by pauses between one and five minutes per participant preference. Each experimental run contained three trial blocks of 5 trials, randomized in order, corresponding to the three experimental conditions. The haptic device was recalibrated between runs. The calibration procedure involved placing the device's stylus inside its inkwell, which allowed its software to match physical and virtual probe positions, which could eventually become incongruent through usage and cause inconsistent sensory feedback.  \autoref{fig:procedure} illustrates the study procedure.
\subsection{Apparatus}
The hardware setup included a custom desktop computer (Intel Core i9-9900KF CPU 3.60GHz, 32 GB RAM, NVIDIA GeForce RTX 2080 Ti) with a configuration for stable VR and haptic rendering at appropriate refresh rates. The application was developed on Windows 10 using Unity 2021.3.5f1 with the OpenHaptics and SteamVR plugins. The virtual scene presenting the stimuli was visually rendered using a Valve Index VR headset, which is a high-resolution device with low latency, precise tracking, a high refresh rate, and a wide field of view. These were essential for the appropriate rendering of 3D stimuli. Haptic interaction utilized 3D Systems Touch grounded force-feedback device. Such devices are commonly utilized in similar studies due to their proven capacity to convey intrinsic object information \cite{marriott_multisensory_2018}. 

\begin{figure}[!ht]
\centering
\includegraphics[width=\linewidth]{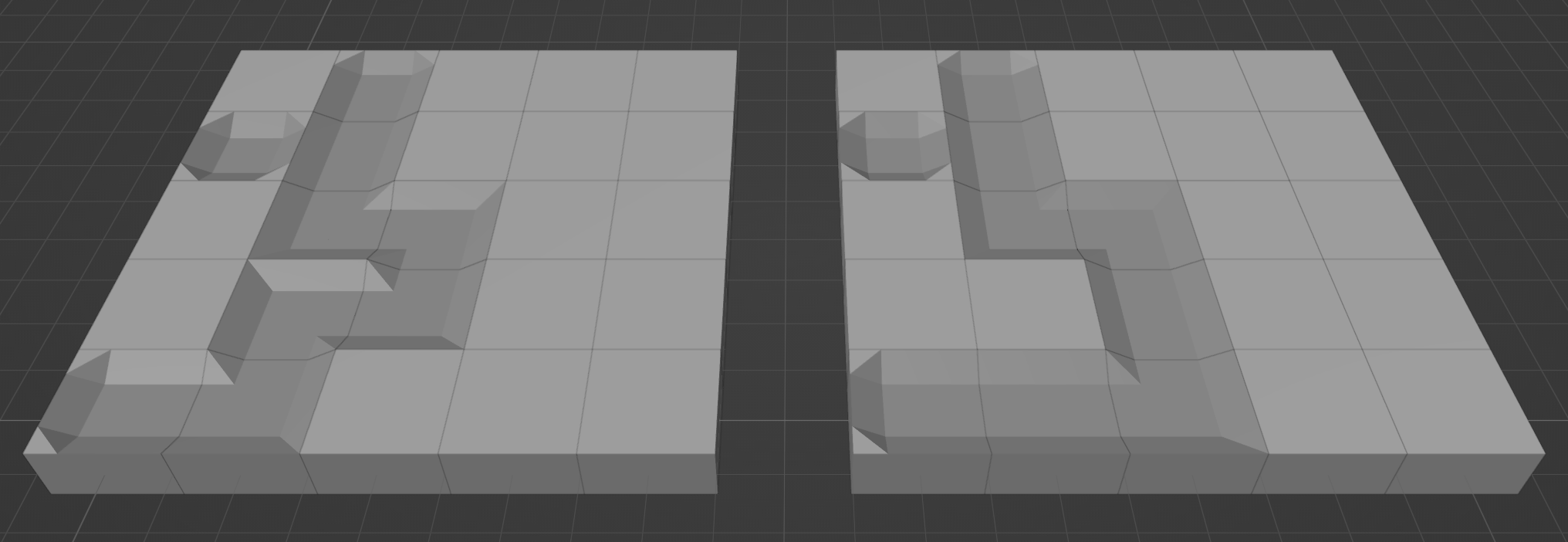}
\caption{Sample stimulus \textbf{(left)} and its corresponding Foil stimulus \textbf{(right)} used in a 2AFC. The stimuli are 3D 5x5 matrices with continuous paths of 8 connected beveled blocks and one beveled block outside the path. The foil stimulus differs from the sample by a one-block change either to the path or the single modified block. Each trial presented a unique Sample/Foil pair for each trial.} \label{stimulus_inclined}
\end{figure}
\subsection{Stimuli}
We created a stimulus with a unique shape for each trial to avoid any stimulus-driven confounds across trials and conditions. The same stimulus set was used across participants, where trials were randomized in order. \autoref{stimulus_inclined} (left) shows an example stimulus. The stimuli comprised square surface patches composed of beveled blocks on 5x5 matrices, inspired by the design presented by Phillips and Christie \cite{phillips_interference_1977} and further developed to fit three-dimensional requirements as in Cattaneo et al. \cite{cattaneo_supramodality_2008}. However, differently from the latter, our stimuli presented height variations instead of changes in color and texture as these are modality-specific object information \cite{lacey_vision_2007}. To generate  the stimuli, we employed a self-avoiding walk algorithm \cite{madras_pivot_1988} to connect target positions, randomly placing eight blocks as continuous paths that visited selected cells exactly once. After creating a path, the algorithm randomly modified one block outside the path. Modified cells were beveled by 1/4 of their respective height, with each side containing a 45-degree slope that allowed the probe to unhinderedly slide in and out of the stimulus path. A beveled path was chosen, as we expected participants to employ \textit{contour following} to explore stimuli within the limited encoding time \cite{lederman_hand_1987}. Beveling adapted to the cell's position within the path, whether it functioned as an extremity, corner, or pass-through path block. We controlled for differences in size and color to prevent \textit{modality encoding bias} that could aid identification in a single modality \cite{lederman_cognitive_1996}. For every stimulus, a foil stimulus (see~\autoref{stimulus_inclined}; right) was created to be presented with the stimulus in the 2AFC task described below. Foils differed from the original stimulus by a one-cell change to either the stimulus' path or to its single modified cell. For path modifications, beveled slopes changed to keep the path as continuous as on its accompanying sample. When single cells were targeted, they changed location to a random neighboring available cell. 

\subsection{Task}
\begin{figure*}
\centering
\includegraphics[width=\linewidth]{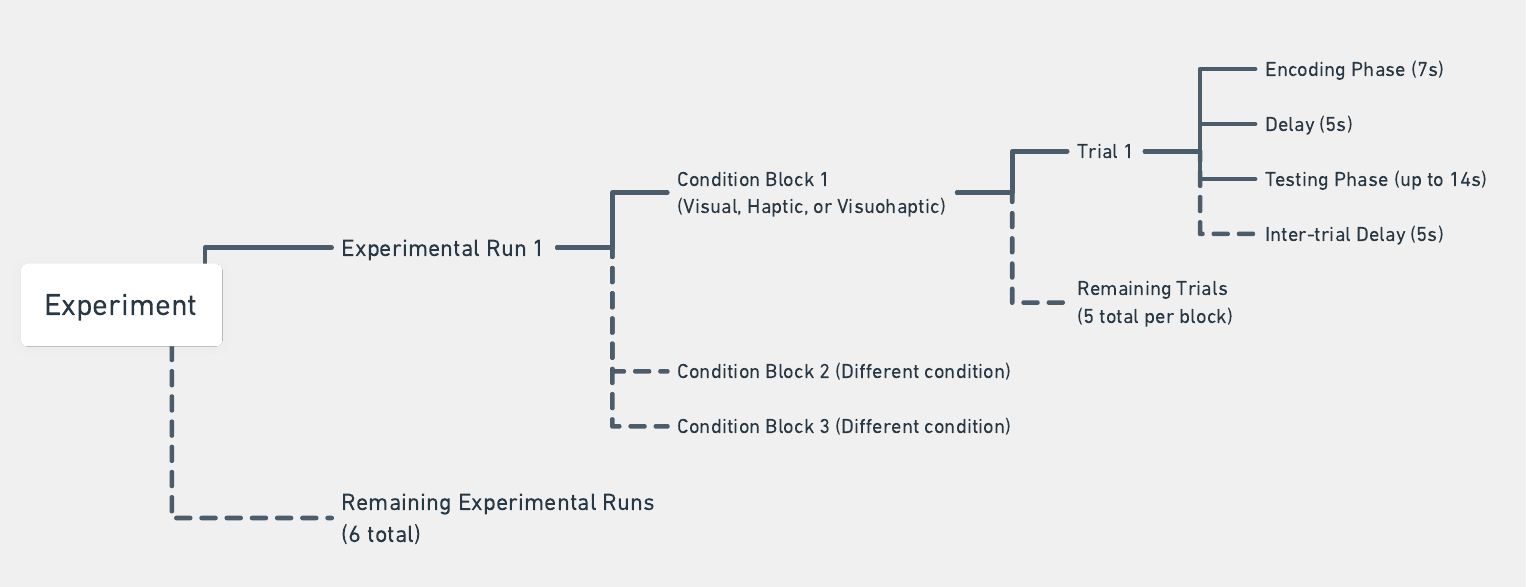}
\caption{Summary of the experiment. Six experimental runs were divided into three counterbalanced single-condition blocks, each containing five trials of its condition. Each trial was divided into: 1) the encoding phase, 7 seconds; 2) the delay: 5 seconds; and 3) the testing phase: up to 14 seconds.} 
\label{diagram}
\end{figure*}

\autoref{diagram} summarizes the task structure. We administered a \textit{DMTS task}, a cognitive assessment tool involving the presentation of a sample stimulus for encoding, followed by a delay phase without the presence of the stimulus to maintain a mental representation of it \cite{daniel_delayed_2016, miller_plans_1968}. At the end of the delay, participants performed a \textit{2AFC} task, in which they had to identify the retained sample alongside a \textit{foil} - a distractor stimulus resembling the original sample \cite{bogacz_physics_2006}. In the stimulus encoding phase, a sample stimulus of one of the three experimental conditions (i.e., visual, haptic, or visuohaptic) was presented at the center of the working space for seven seconds. After a five-second delay, participants were allowed up to 14 seconds to explore sample and foil stimuli to report which was identical to the retained stimulus. To indicate their decision in this 2AFC task, participants used the left-hand thumb VR controller joystick to reveal the side of the target stimulus. The target positions were random and balanced across trials to exclude the effects of potential response biases. Upon responding, participants received visual feedback regarding the correctness of their choice, either as a green thumbs-up or a red thumbs-down, displayed in the middle of the scene. A five-second inter-trial interval separated answering from the start of the subsequent trial. Participants were asked to raise the haptic device's stylus to keep it from touching the stimuli cover before the beginning of a new trial.  

\begin{figure}[!ht]
\centering
\includegraphics[width=\linewidth]{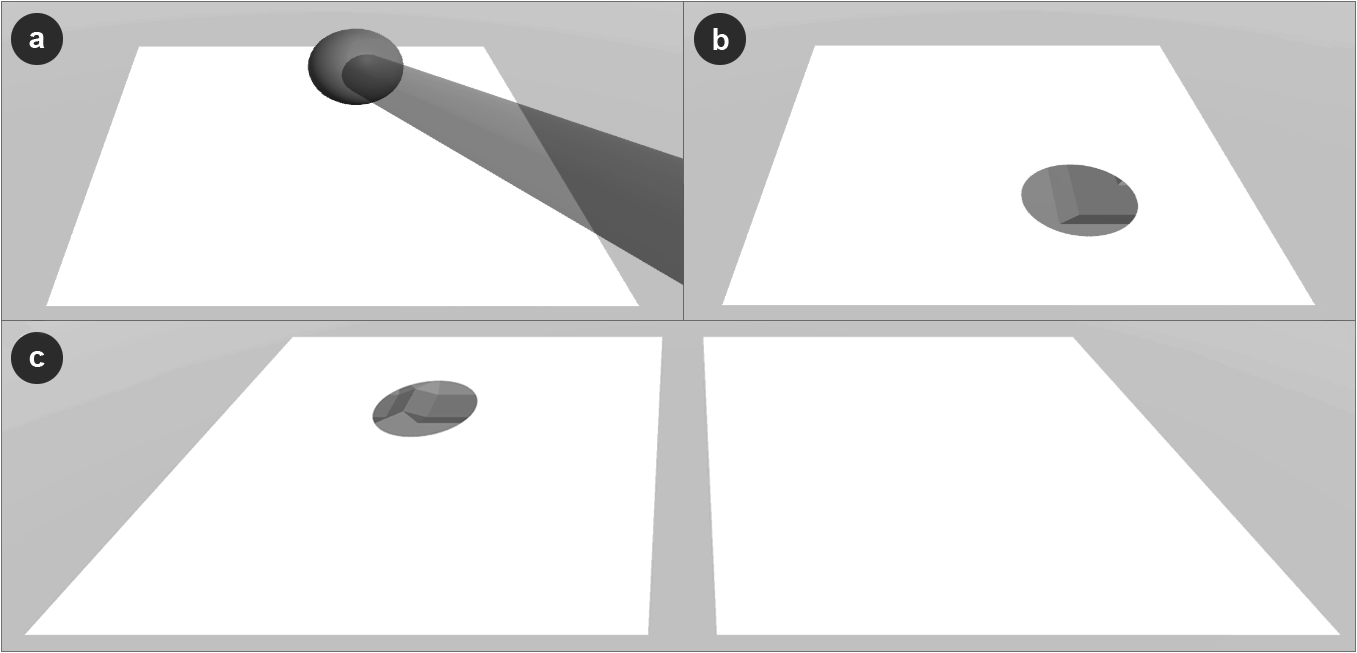}
\caption{\textbf{(a):} A participant moves the virtual probe towards a sample during the learning phase. \textbf{(b):} A mask reveals the stimulus at the touched location. \textbf{(c):} Sample and foil exploration during the testing phase.} \label{trial_collage}
\end{figure}

As illustrated in \autoref{trial_collage}, a gray layer covered stimuli in all conditions in both the learning and testing phases. The stimuli could only be explored using the probe to perform \textit{exploratory procedures}, such as lateral motion and contour following \cite{lederman_hand_1987}, that would reveal the stimuli through a circular aperture window at the touched location. This sampling limitation aimed to make the different experimental conditions comparable, as vision and touch significantly differ in the rate and range in which they can encode object information, with vision being able to sample at a fraction of the time it would take touch \cite{lederman_visual_1990, lederman_hand_1987}. Haptic exploration is typically performed through successive impressions \cite{gibson_observations_1962}, but as our task design limited visualization to only the touched areas, both visual and haptic exploration are to be performed at the approximately same pace and scope of information sampled per time. Additionally, the task was designed in such a manner as to cause participants to explore stimuli through similar movements under the three conditions as to prevent a confounding factor, as humans generally rely on vision and prioritize it to perceive geometric features \cite{zangaladze_involvement_1999}. Our choice for viewport diameter was inspired by results from Loomis, Klatzky, and Lederman, who found comparable performance between unimodal haptic and vision conditions when the simulated visualization window's aperture was equivalent to the touched area \cite{loomis_similarity_1991}. Participants could reveal the stimulus's appearance at the probe location in the visual condition. Under the visuohaptic condition, participants could examine the properties of stimuli through both vision and touch. In contrast, under the haptic condition, participants could only obtain haptic information at touched locations without visual cues.
\subsection{Independent Variables}
We define \textsc{encoding modality} as an independent variable in our study design. The \textsc{Encoding Modality} contains three conditions: \textit{Visual}, \textit{Haptic}, and \textit{Visuohaptic}. In trials of the \textit{Visual} condition, participants explored the sample stimulus only visually through an aperture window, as described above. The haptic feedback was limited to a flat surface to assert that they were in contact with the stimulus, but no force feedback on the stimulus characteristics was applied. In contrast, force feedback was provided in trials in the \textit{Haptic} condition, while no visual stimulus characteristics were presented. In the \textit{Visuohaptic} condition, both force feedback and visual appearance were presented.

\subsection{Dependent Variables}
We measure \textit{Error Rate} and \textit{Response Time} as dependent variables. \textit{Error Rate} is assessed as the relative number of incorrect responses, namely selecting the foil instead of the target stimulus. \textit{Response Time} was evaluated as the elapsed time between the onset of the testing phase, i.e., the onset of target and foil stimulus display, and response action, i.e., the completion of a lateral joystick movement.

\section{Results}
A total of twenty-three participants completed the study. From the initial pool of participants, those whose overall performance was below 60\% (fewer than 54 out of 90 correct responses) were excluded from analysis, as this cutoff represents the minimum score required to be performing significantly above chance at $p < .05$ according to a binomial probability distribution. Thus, the final analysis included a total of 20 participants. We statistically investigate the error rate and the response time for significant differences between the conditions. We use a Shapiro-Wilk test to examine the distribution of our measures. Greenhouse–Geisser corrections are applied whenever the assumption of sphericity is violated. Then, we apply statistical testing to investigate our measures for statistical differences. We report Cohen's d as a measure for the effect size.

\begin{figure}[!ht]
\centering
  \includegraphics[width=\linewidth]{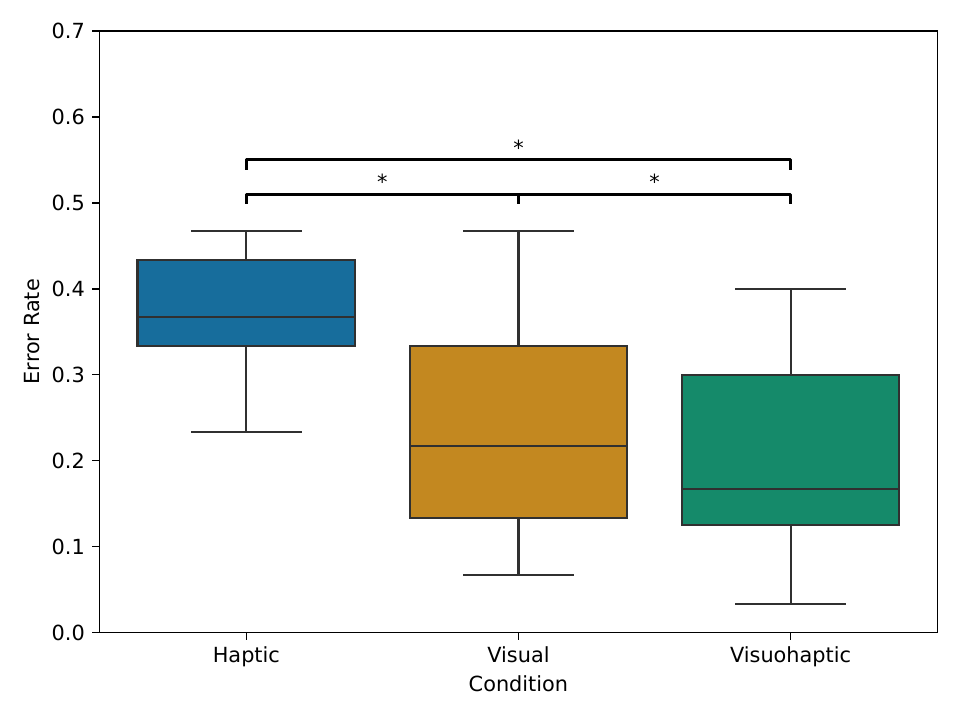}
  \caption{Error rates per condition. Asterisks indicate significant differences between conditions, as determined by Bonferroni-corrected t-tests. Significant differences were found between all conditions. Trials in the visuohaptic condition resulted in the lowest error rates, while trials in the haptic-only condition resulted in the highest error rates.}
  \label{fig:num_errors}
\end{figure}

\subsection{Error Rate}

A Shapiro-Wilk test indicated that the error rate was normally distributed ($p > .05$). Thus, we conducted a repeated measures ANOVA, revealing a significant main effect for our conditions, $F(2, 38) = 30.59, p < .001$. Bonferroni-corrected post-hoc t-tests revealed a substantial difference between haptic and visual, $t(19) = 4.73, p < .001, d = 1.27$, haptic and visuohaptic, $t(19) = 6.75, p < .001, d = 1.93$, and visual and visuohaptic conditions, $t(19) = 3.3, p = .011, d = 0.41$. The haptic-only condition elicited the highest error rate (M = 0.37, SD = 0.06), followed by visual (M =0.24, SD = 0.13) and visuohaptic conditions (M = 0.19, SD = 0.11). \autoref{fig:num_errors} shows the mean error rate per participant.

\begin{figure}[!ht]
  \centering
  \includegraphics[width=\linewidth]{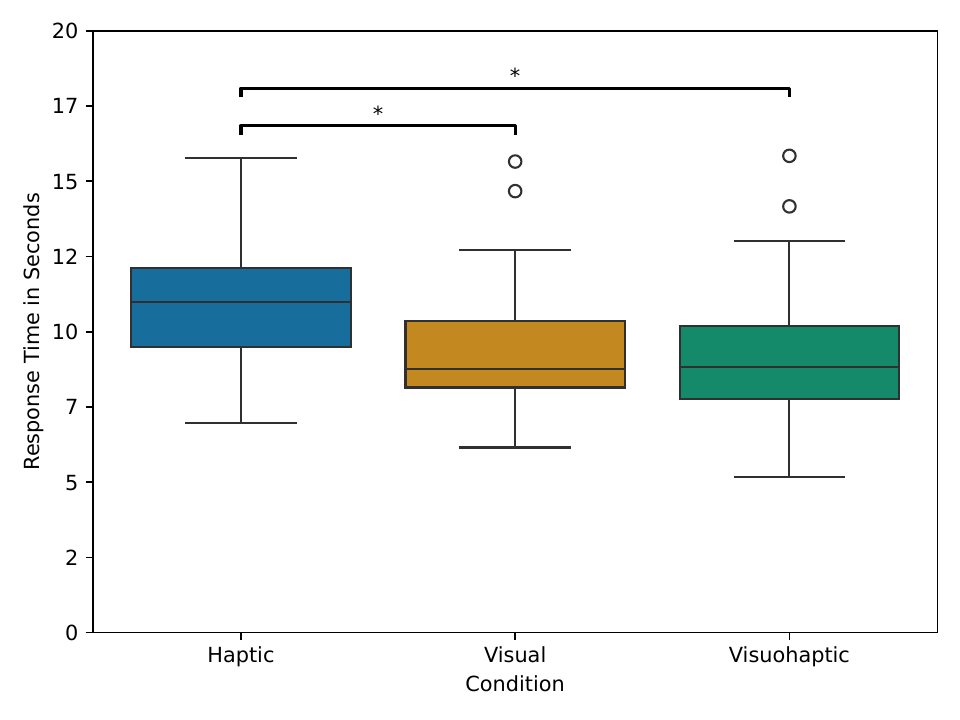}
\caption{Averaged response times per condition. Asterisks indicate significant differences between conditions, as determined by Bonferroni-corrected Wilcoxon signed-rank tests. Haptic-only encoding resulted in the highest response times, while both visual and visuohaptic encoding resulted in lower response times.}
\label{fig:tct}
\end{figure}
\subsection{Response Time}
A Shapiro-Wilk test indicated that response times were not normally distributed ($p < .05$). We, therefore, applied a Friedman test for non-parametric testing. The Friedman test resulted in a significant main effect, $\chi^2(2) = 10.9, p = .004$. Bonferroni-corrected Wilcoxon signed-rank post-hoc tests revealed a significant difference between haptic and visual encoding, $p = .012, d = 0.63$, and haptic and visuohaptic encoding, $p = .003, d = 0.68$. No significant difference was found between visual and visuohaptic encoding ($p > .05$). Haptic encoding caused the highest  response times (M = 11.02, SD = 2.17), followed by visual (M = 9.50, SD = 2.58) and visuohaptic encoding (M = 9.39, SD = 2.57). \autoref{fig:tct} illustrates the response times.

\section{Discussion}
Here, we investigated the effects of supplementing VR visualization with haptic force feedback. Our results showed that integrating haptic information during stimulus encoding (visuohaptic condition) lowered error rates compared to visual encoding alone, supporting our main hypothesis (H1). We did not find a reduction in response time (H2). Together, these results show the potential benefits of forming a more accurate mental representation of an object if haptic force feedback is provided in addition to visual information and may inspire future research to improve the performance in different tasks that rely on an accurate representation of a 3D object. 

\subsection{Visuohaptic Encoding Reduces Error Rates}
The reduced error rates that we observed in our results are in line with previous research concerning the integration of sensory information from different modalities into shared representations \cite{engelkamp_human_1994, loomis_analysis_1982, klatzky_identifying_1985, gaissert_categorizing_2012}. In the visuohaptic encoding condition, the integration of visual and haptic sensory input can provide redundant and complementary information about the same stimulus, which may lead to a more accurate mental representation of three-dimensional objects \cite{engelkamp_human_1994, newell_viewpoint_2001}. Visuohaptic encoding may also have facilitated object identification by providing more retrieval cues than either visual or haptic unimodal encoding alone \cite{engelkamp_visual_1995}. \textit{Multisensory enhancement} might also elucidate the observed effect, as the response to bimodal stimulation yielded greater accuracy than the response to vision, the most effective of its component senses \cite{stein_multisensory_2008}. Another potential account for the observed advantage of visuohaptic encoding would be the \textit{early multisensory facilitation} process described by Stein and Meredith \cite{stein_merging_1993}, which posits that responses to multiple congruent sensory information about objects may underlie behavioral benefits. Edelman explains this phenomenon through the concept of \textit{reentry}, the explicit interrelating of different simultaneous representations across sensory modalities that mutually educate their counterparts and result in enhanced learning \cite{edelman_neural_1987, smith_development_2005}. 

Although our experiment is not directly comparable with previous studies, our results are generally aligned with their observations in that visuohaptic integration reduces error rates. Kreimeier et al. also observed lower error rates in visuohaptic object identification, although comparability is limited as their corresponding experiment employed exoskeleton gloves and a limited number of familiar primitive shapes, whereas our experiment utilized grounded force-feedback and a much higher number of unfamiliar and more complex objects \cite{kreimeier_evaluation_2019}. Although their experiment explored a different paradigm, Jüttner et al. also observed significant effects of haptic integration on object identification accuracy \cite{juttner_when_2001}. Reporting on a pattern-matching task involving a fingertip interface, Seaborn et al. reported that dual-coding of both visual and haptic modes increases recall accuracy \cite{seaborn_exploring_2010}. Performance-hindering effects of visuohaptic integration were reported by Stahlman and colleagues, who attributed the lower performance of visuohaptic encoding to overshadowing. However, it is possible that manual exploration of physical objects impacted the visual component of bimodal encoding as it may have occluded viewing, which would explain why haptic training disrupted visual recognition at test while the opposite effect was not observed \cite{stahlman_overshadowing_2018}. Desmarais et al. presented similar error rates when comparing bimodal encoding with haptic and visual conditions \cite{desmarais_visuo-haptic_2017}. However, the experiments they described are arguably different in both encoding and testing settings as their bimodal encoding of physical objects occurred sequentially, while the testing phases in their experiments involved separate haptic and visual phases as they were interested in the effects of cross-modality. In summary, our findings are generally consistent with the trend observed in previous studies of different designs and scopes, with a few exceptions that may be attributed to stark differences in experimental paradigms and encoding conditions. This suggests that visuohaptic integration can enhance object identification accuracy.

\subsection{Visuohaptic Encoding Did Not Reduce Response Time}
Differently from what we hypothesized, our results did not support a significant difference between visuohaptic encoding and its visual counterpart in terms of response latency, which may be a surprising outcome given the accompanying error rate results. Authors who observed similar effects, such as Miquée et al., have provided a rationale for such a discrepancy and described its potential causes, noting that in their object discrimination task using position tracking and neuroimaging, participants only expressed their decisions after thoroughly exploring both target and foil objects instead of responding as soon as salient and decisive differences between these objects were detected \cite{miquee_neuronal_2008}. The unexpected discrepancy between observed accuracy and response latency performance scores might also involve item-level feedback, which in our experiment included answer correctness but not response times, potentially causing participants to shift their position on their speed-accuracy tradeoff curve \cite{ackerman_exploiting_1999} and prioritize accuracy over speed performance \cite{kyllonen_use_2016}. The dynamics of the task's explicit time limit and the two-alternative forced choice might also account for unexpected response latency results as participants might make assumptions about the optimal decision time based on the response deadline and be inclined to approach the upper limits \cite{bogacz_physics_2006}, especially if participants perceive a higher incentive for accuracy than for speed.

Our results align with findings from Seaborn et al., who did not encounter a significant difference in participants’ task times between the visual and visuohaptic modes \cite{seaborn_exploring_2010}. Desmarais et al. even reported slower response times for participants in the visuohaptic learning condition in both visual and haptic identification modalities \cite{desmarais_visuo-haptic_2017}. Accordingly, Kreimeier et al. reported slower response times for the visuohaptic condition in comparison with its unimodal visual counterpart. However, the authors admit potential performance-deterring inadequacies of exoskeleton force feedback gloves used in their apparatus, as such gloves are still in the early phases of technological development in comparison to other haptic displays \cite{kreimeier_evaluation_2019}. 

\subsection{Performance on the Haptic Condition Contrasts with Visual and Visuohaptic Counterparts}
As proof of principle and replication of a highly expected effect, the large performance differences observed between the unimodal haptic encoding and conditions comprising a visual component were in line with expectations. This finding confirms that our experimental design effectively captured the well-established phenomenon that humans generally rely on vision and prioritize it to perceive and compare spatial characteristics of unfamiliar objects \cite{rock_vision_1964,zangaladze_involvement_1999, lacey_mental_2006}. In comparing visual and haptic systems, Lederman et al. stated that haptics produce substantial error when deriving information about spatial properties of objects and attributed this trend to the system's low spatial resolution as well as its memory and integration demands \cite{lederman_visual_1990}. Adequately extracting object characteristics solely through haptic exploration occurs at a much slower rate than vision \cite{liu_haptic_2009}. Newell et al. needed to allow participants more time for haptic learning in comparison to its visual counterpart to equate these encoding modalities \cite{newell_viewpoint_2001}, so it was plausible to expect our experiment's haptic condition to underperform as it was given the same amount of time as the visual condition. Lacey and Sathian attribute this asymmetry between vision and touch to their relative efficiencies in encoding shape information, which might be affected by competition from other salient modality-specific object properties such as hardness and texture \cite{lacey_visuo-haptic_2014}. 

Our performance levels for this encoding condition are similar to within-modality object identification numbers reported by Klatzky and Ernst \cite{klatzky_identifying_1985, ernst_multisensory_2007}. Cross-modality studies also generally agree on the superiority of visual encoding over its haptic counterpart. \cite{walk_developmental_1981, lewkowicz_development_1994, lacey_mental_2006}. Among comparable studies, a similar gap between haptic and visual conditions was observed by Seaborn et al. \cite{seaborn_exploring_2010}. Stahlman et al. also reported lower accuracy performance in the haptic encoding condition, although their testing was unimodal \cite{stahlman_overshadowing_2018}. In a matrices task that enforced sequential exploration through a circular aperture window similarly to our experiment, Cattaneo and Vecchi found that visual stimuli performed better than their haptic counterparts \cite{cattaneo_supramodality_2008}. Desmarais et al. observed similar results for error rates but not for response latency, which may be explained by differences in experimental design, such as the use of physical objects, the choice for unimodal instead of bimodal for the testing phase, and the fact that participants in the haptic condition were granted more than double the encoding time in comparison with its visual counterpart \cite{desmarais_visuo-haptic_2017}. 

Although our study did not aim to compare visual and haptic or visuohaptic and haptic conditions, we included the unimodal haptic condition as a control to validate the appropriateness of our unimodal visual condition. Our experiment's visual condition could have been affected by experimental and stimulus design choices, such as constraining stimulus visualization to probed areas, even though this visual constraint resembled the established procedures described by Loomis et al. \cite{loomis_similarity_1991} and Cattaneo and Vecchi \cite{cattaneo_supramodality_2008}. The fact that the unimodal haptic condition did not outperform its visual counterpart and that the performance gap between these conditions resembled previous findings indicated that the visual condition's presentation did not deteriorate due to our experimental design. Thus, results for the haptic condition and its proportionality to the visual condition support the validity of comparisons between visuohaptic and visual encoding conditions, thus supporting H2.

\subsection{An Experimental Framework for Comparing the Effects of Encoding Modalities on Representations}
Our study contributes an experimental framework that operationalizes a working memory task that estimates the accuracy of object representations. Forming a mental representation of an object is the first step for any task that you then do with the said object. If mental representations are formed with high accuracy, then any decision-making, comparison, and object manipulation task will possibly be performed with higher precision. The most direct experimental way to test for the accuracy of a mental representation is to test it in a DMTS task. This is because it allows us to measure the accuracy of a judgment that requires detailed and accurate knowledge/mental representation of an object. Therefore, such a task is susceptible to assess differences in how well a modality-overarching mental object representation has been formed.  Additionally, we address the challenges researchers face in isolating and comparing the effects of encoding modalities on the retention of object representations. For example, Desmarais et al. reported limitations in assessing visuohaptic integration effects as reliable visual information in their task eliminated the need to acquire haptic cues \cite{desmarais_visuo-haptic_2017}. Humans are biased towards encoding novel object information through vision, which can simultaneously probe different areas of an object. In contrast, haptics is generally limited to sampling information through successive impressions \cite{kassuba_vision_2013, woods_visual_2004}. Our design homogenizes these modalities by constraining visual presentation to areas that participants are actively probing, similar in different aspects to the designs described by Loomis~\cite{loomis_similarity_1991} and Cattaneo~\cite{cattaneo_supramodality_2008}. Another common challenge to the investigation of visuohaptic integration is the occlusion of visual cues during haptic exploration, which may potentially be responsible for asymmetric interferences in cross-modal object identification \cite{stahlman_overshadowing_2018}. Although other researchers have succeeded at avoiding partial occlusion by presenting visual and haptic stimuli sequentially \cite{desmarais_visuo-haptic_2017}, the temporal congruence between bimodal presentations is known to affect working memory performance \cite{ghazi_bimodal_2022}. As our experiment leverages VR, hiding the virtual probe for the duration of its contact with stimuli enables participants to leverage unobstructed  viewing of probed areas and synchronous haptic rendering in the visuohaptic condition. Another critical challenge tackled by our design's homogenization of sensory conditions is the equalizing of exploratory movements, which is critical as active stimulus exploration enhances performance \cite{harman_active_1999}. Our design levels hand motion across conditions to prevent the \textit{enactment effect} \cite{engelkamp_visual_1995}, as \textit{subject-performed tasks} would be prevalent in the unimodal haptic-only exploration as the sole condition requiring active motor exploration\cite{gibson_observations_1962}, whereas fully-exposed visual stimuli would require only passive encoding \cite{brooks_specificity_1999}. Making exploratory movements comparable across conditions is paramount to prevent a potential confounding factor, as subject-performed tasks tend to generate higher object information recall \cite{engelkamp_human_1994}.

\subsection{Implications for the Design of Data Visualization Interfaces}

Our study's findings have practical implications for improving the design of interactive data visualization interfaces and their ability to promote object information retention as representations created through visuohaptic learning might be robust and, therefore, easier to recall than their visual counterparts. As we observed that visuohaptic integration yields lower error rates than visual encoding, incorporating haptic feedback into immersive visualizations may help users memorize details of complex objects. The benefits of visuohaptic integration might be particularly relevant to professionals, such as paleontologists and surgeons, who rely on accurate mental representations of digital objects to carry out their procedures \cite{cunningham_virtual_2014, meier_virtual_2001, siqueira_rodrigues_design_2023}. Previous research on immersive visualization design spaces recommended leveraging haptics as an encoding channel in multi-sensory presentation, as virtual reality enables egocentric perception of spatialized data across senses \cite{saffo_unraveling_2024, moloney_affordance_2018}. Our findings align with such insights, suggesting that integrating haptics into immersive visualizations may improve memorability. Nevertheless, it is essential to note that our findings represent an initial step in assessing the potential advantages of incorporating haptics in the data visualization context, as further work is necessary to address the limitations of our experimental design. Our study employed a DMTS paradigm with a delay that falls within a short-term memory timeframe \cite{daniel_delayed_2016}. Thus, our findings may have higher applicability to tasks that require maintaining the characteristics of 3D objects to perform a task with it. In contrast, effects on long-term memory have to be explored in future studies. Image segmentation, for example, requires professionals to perform selections from given viewpoints while maintaining coherent representations of that object as memorized from other viewpoints \cite{sanandaji_developing_2023}. Similar short-term memorization is necessary for 3D modeling, where practitioners must mentally rotate object representations to predict the outcomes of given actions while designing 3D objects \cite{kadam_improvement_2012}. Further research is also needed to improve the ecological validity of our findings in assisting specific professional workflows and their corresponding digital objects. Our experiment employed synthesized stimuli to control complexity as a means to avoid creating confounding factors. However, as complexity remained constant, we cannot assert that effects would be observable under different levels of complexity. Our synthetic stimuli were also specifically conceived to enable unobstructed exploration of samples, which was essential to enable participants to complete trials in a short amount of time so that our study could have enough instances for inferential statistics. However, such a study design choice limits the generalizability of our results. Future research might benefit from testing a similar study design using real-world unaltered stimuli, which could further validate our findings towards specific scenarios or increase the generalizability of our findings.

\subsection{Limitations and Future Work}
The first limitation of our study is its relatively small sample size of participants of specific ages and handedness. Twenty participants remained in our analyses after three datasets were excluded as participants were either unable or unwilling to perform the task and did, therefore, not exceed chance-level performance. A post-hoc power analysis indicated that, given the observed effect size between visual and visuohaptic conditions, a sample size nearly five times larger would be necessary to reach conventional statistical power levels. While this constraint is common in HCI \cite{caine_local_2016}, this limitation impacts the robustness of our findings. Thus, future studies with larger sample sizes and variations of the task and stimulus material will contribute to the generalizability of the reported effects. Participation was limited to those declaring not having had experience with grounded force-feedback devices. While this restriction was important to prevent confounding factors and decrease between-subject variability, the adjustment to using the device might vary between participants. As prior research indicates that performance with haptic devices generally improves through short practice \cite{goos_can_2001}, future research could benefit from testing the utility of haptic feedback integration in longitudinal study designs. Additionally, we utilized a 6-degree-of-freedom (DOF) grounded force-feedback device. Our findings' generalizability should be further tested with other haptic feedback devices, e.g., devices with different DOFs, vibrotactile displays, or exoskeleton data gloves. As we used relatively simple force feedback, future studies should test for the effects of supplementing other haptic information such as viscosity, vibration, and friction. Another limitation of our experiment is that it did not control for potential interference of verbal strategies, which Lacey and Campbell claim to ``facilitate encoding of unfamiliar objects regardless of modality'' \cite{lacey_mental_2006}. While verbalization could equally affect visual and visuohaptic conditions, further research could aim to suppress verbal rehearsal strategies \cite{baddeley_exploring_1984} or test for verbal interference \cite{baddeley_working_1974}. Since the literature establishes that VR impacts object memorability, which may impact other factors such as mental workload~\cite{10.1145/3582272}. Previous research suggested that integrating psychophysiological measures in immersive settings will provide real-time insights into user perception~\cite{9756768}. Consequently, the integration of electrodermal activity~\cite{10.1145/3319499.3328230, 10.1145/3604270} or electroencephalography~\cite{10.1145/3313831.3376766} has been suggested by previous work. Finally, as several variables can affect cognitive performance in VR \cite{laviola_3d_2017}, it could be valuable to replicate our experimental design using non-immersive rendering and compare our results to better attribute memorability effects to haptic technology. Since immersive visualization carries increased costs to both system designers and users, assessing whether desktop-based applications could benefit from the memorability effects of visuohaptic integration could broaden the impact of our findings.

\section{Conclusion}

Our findings suggest that integrating haptic feedback into immersive encoding of 3D objects may enhance memory accuracy. This finding indicates that visuohaptic encoding could form more accurate or detailed mental representations of 3D objects than its vision-only counterpart. Our contribution extends the existing literature by introducing a delayed match-to-sample task that homogenized information sampling across modalities and confirmed its appropriateness through a haptic-only control condition, which allowed us to attribute the reported behavioral benefits to the integration of haptic information. Altogether, our findings have implications for the design of interactive systems aimed to convey the characteristics of complex digital objects in synergy with the human capacity to integrate sensory information optimally. Interface designers might consider leveraging visuohaptic integration in immersive data exploration to better serve professionals whose workflows require accurately recalling object characteristics.

\section*{Data Availability}
\label{sec:data-availability}
The data that support the findings of this experiment, along with their corresponding data analysis scripts and software on GitHub\footnote{\url{https://github.com/lsrodri/VHMatch}}.

\begin{acks}
The author acknowledges the support of the Cluster of Excellence »Matters of Activity. Image Space Material« funded by the Deutsche Forschungsgemeinschaft (DFG, German Research Foundation) under Germany's Excellence Strategy – EXC 2025 – 390648296.
\end{acks}

\bibliographystyle{ACM-Reference-Format}
\bibliography{main}










\end{document}